\documentstyle[preprint,aps]{revtex}

\draft
\newcommand{\be}{\begin{equation}}
\newcommand{\ee}{\end{equation}}
\newcommand{\ben}{\begin{eqnarray}}
\newcommand{\een}{\end{eqnarray}}

\newcommand{\iii}{\'{\i}}
\newcommand{\nn}{\nonumber}

\begin{document}

\draft
\title{Inclusion relations among separability criteria}
\author{J. Batle$^1$, A. R. Plastino$^{1,\,2,\,3}$, M. Casas$^1$, and A.
Plastino$^{2}$}

\address {$^1$Departament de F\iii sica, Universitat de les Illes Balears,
07071 Palma de Mallorca, Spain \\ $^2$National University La Plata and
Argentina's National Research Council (CONICET), C.C. 727, 1900 La Plata,
Argentina\\ $^3$ Department of Physics, University of Pretoria, 0002 Pretoria,
South Africa}


\maketitle
 \begin{abstract}

 We revisit the application of
different separability criteria by recourse to an exhaustive  Monte Carlo
exploration involving the pertinent state-space of pure and mixed states. The
corresponding  chain of implications of different criteria is in such a way
numerically elucidated. We also quantify, for a bipartite system of arbitrary
dimension, the proportion of states $\rho$ that can be  distilled according to
a definite criterion. Our work can be regarded as a complement to the recent
review paper by B. Terhal [Theor. Comp. Sci. {\bf 287} (2002) 313]. Some
questions posed there receive an answer here.

\noindent
 Pacs: 03.67.-a; 89.70.+c; 03.65.Bz


\end{abstract}

\maketitle

\newpage

\section{Introduction}

The development of criteria for entanglement and separability is one aspect of
the current research efforts in quantum information theory that is receiving,
and certainly deserves, considerable attention \cite{T02}. Indeed, much
progress has recently been made in consolidating such a cornerstone of the
theory of quantum entanglement \cite{T02}. The relevant state-space here is of
a high dimensionality, already 15 dimensions in the simplest instance of
two-qubit systems. The systematic exploration of these spaces can provide us
with valuable insight into some of the theoretical questions extant.

 As a matter of fact, important steps have been recently made towards a 
systematic exploration of the space of arbitrary (pure or mixed) states of 
composite quantum systems \cite{ZHS98,Z99,Z01} in order to determine the typical 
features exhibited by these states with regards to the phenomenon of quantum
entanglement \cite{ZHS98,Z99,Z01,MJWK01,IH00,BCPP02a,BCPP02b}. Entanglement is
one of the most fundamental and non-classical features exhibited by quantum
systems \cite{LPS98}, that lies at the basis of some of the most important
processes studied by quantum information theory \cite{LPS98,WC97,W98,NC00,GD02}
such as quantum cryptographic key distribution \cite{E91}, quantum
teleportation \cite{BBCJPW93}, superdense coding \cite{BW93}, and quantum
computation \cite{EJ96,BDMT98}.

It is well known \cite{T02} that,  for a  composite quantum system, a state
described by the density matrix $\rho$ is called ``entangled" if it can not be
represented as a mixture of factorizable pure states. Otherwise, the state is
called separable. The above definition is physically meaningful because
entangled states (unlike separable states) cannot be prepared locally by acting
on each subsystem individually \cite{P93}.

The separability question has  quite interesting  echoes in information theory
and its associate information measures or entropies. When one deals with a 
classical composite system described by a suitable probability distribution 
defined over the concomitant phase space, the entropy of any of its subsystems 
is always equal or smaller than the entropy characterizing the whole system. 
This is also the case for separable states of a composite quantum
system \cite{NK01,VW02}. In contrast, a subsystem of a quantum system described 
by an entangled state may have an entropy greater than the entropy of
 the whole system. Indeed, the von Neumann entropy of either
 of the subsystems of a bipartite quantum system described (as a whole)
 by a pure state provides a natural measure of the amount of entanglement
 of such state. Thus, a pure state (which has vanishing entropy)
 is entangled if and only if its subsystems have an entropy
 larger than the one associated with the system as a whole.

  Regrettably  enough, the situation is more complex when the composite system
  is described by a mixed state: there are
  entangled  mixed states such that the entropy of the complete
  system is smaller than the entropy of one of its subsystems.
  Alas, entangled mixed states such that the entropy of the
  system as a whole is larger than the entropy of either of
  its subsystems do exist as well. Consequently, the classical
  inequalities relating the entropy of the whole system with the
  entropy of its subsystems provide only necessary, but not
  sufficient, conditions for quantum separability.
  There are several entropic (or information) measures that can be
  used in order to implement these  criteria for separability.
  Considerable attention has been paid, in this regard, to the
  $q$-entropies \cite{T02,VW02,HHH96,HH96,CA97,V99,TLB01,TLP01,A02},
  which incorporate both R\'enyi's \cite{BS93} and Tsallis' 
\cite{T88,LV98,LSP01} families of
  information measures as special instances (both admitting, in turn,
  Shannon's measure as the particular case associated with the limit
  $q\rightarrow 1$). The reader is referred to  Appendix A for a brief review
     on $q$-entropies.

 The early motivation for the studies reported in
  \cite{VW02,HHH96,HH96,CA97,V99,TLB01,TLP01,A02} was
  the development of practical separability criteria for density matrices.
  The discovery by Peres of the partial transpose criteria, which for
  two-qubits and qubit-qutrit systems turned out to be both necessary
  and sufficient, rendered that original motivation somewhat outmoded.
  In point of fact, it is not possible to find a necessary and sufficient
   criterion for separability based
  solely upon the eigenvalue spectra of the three density matrices
  $\rho_{AB}, \rho_A=Tr_B[\rho_{AB}]$, and $\rho_B=Tr_A[\rho_{AB}]$
  associated with a composite system $A\oplus B$ \cite{NK01}.

  Interesting concepts that revolve around the separability issue have been
developed over the years. A beautiful account is given in Terhal in \cite{T02}.
Among them we find criteria like the so-called Majorization, Reduction and
Positive Partial Transpose (PPT) together with the concept of distillability
\cite{T02}. Quantum entanglement is a fundamental aspect of quantum physics
  that deserves to be investigated in full detail from all possible
  points of view. The chain of implications, and the related inclusion relation,
among the different separability criteria is certainly a vantage point worth
   of detailed scrutiny. It is our purpose here to revisit, with such a goal
   in mind, the separability question by means of an exhaustive Monte Carlo
exploration involving the whole  space of pure and mixed states. Such an effort
should shed some light on the inclusion issues that  interest us here.
 Concrete numerical evidence will thus be provided on the relations among the
separability criteria. We will then be able to quantify, for a bipartite system
of arbitrary dimension, the proportion of states $\rho$ that can be
distilled according to a definite criterion. This numerical exploration could
be viewed as a complement on the review paper by Terhal  \cite{T02}, because
some questions posed by her will receive an answer in this work.


The paper is organized as follows. We sketch  in Section II the different
separability criteria to be investigated and discuss some mathematical and
numerical techniques used in our survey in Section III. Our results are
reported in Section IV, and some conclusions are drawn in Section V. For the
sake of completeness, we include an Appendix on $q$-entropies.

\section{Brief sketch on separability criteria}
   From a historic viewpoint, the first separability criterion is that
of Bell (see \cite{T02} and references therein). For every pure entangled state
there is a Bell inequality that is violated. It is not known, however, whether
in the case of  many entangled mixed states, violations exist. There does exist
a witness for every entangled state though \cite{Horodeckis1996}. It was shown
by Horodecki {\it et al.} that a density matrix $\rho \equiv \rho_{AB}$ is
entangled if and only if there exists an entanglement witness
(a hermitian super-operator $\hat W=\hat W^{\dagger}$) such that

\ben \label{wit} Tr\, \hat W\,\rho &\le& 0, \,\,\,\, {\rm while} \nn \\
 Tr\, \hat W\,\rho &\ge& 0,\,\,{\rm for \,\,all\,\,separable\,\,states}.  \een

A special, but quite important LOCC operational separability criterion,
necessary but not sufficient, is provided by the positive partial transpose
(PPT) one. Let $T$ stand for matrix transposition. The PPT requires that  \be
\label{W}  [\hat 1 \otimes \hat T](\rho) \ge 0. \ee

Another operational criterion is called the {\it reduction} criterion, that
is satisfied, for a given state $\rho \equiv \rho_{AB}$, when both \cite{T02}

\ben \label{reducti} \hat 1 \otimes \rho_B -\rho &\ge&  0\nn \\  
\rho_A \otimes \hat 1-\rho &\ge& 0  .\een

Intuitively, the distillable entanglement is the maximum asymptotic yield of
singleton states that can be obtained,  via LOCC, from a given mixed state.
Horodecki {\it et al.} \cite{Horo97} demonstrated that any entangled mixed state
of two qubits can be distilled to obtain the singleton. This is not true in
general. There are entangled mixed states of two qutrits, for instance, that 
cannot be distilled, so that they are useless for quantum communication.
 In our scenario an important fact is that all states that violate
the reduction criterion are distillable  \cite{Horo99}.

Entanglement witnesses completely characterize the set of separable states.
Alas, they are not usually associated to a simple computational treatment,
except in the PPT instance.  Thus, in order to decide whether a given state
$\rho$ is entangled one needs additional criteria, functional separability ones
\cite{T02}. One of them associates PPT to the rank of a matrix. Consider two
subsystems $A$, $B$ whose description is made, respectively, in the Hilbert
spaces ${\cal H}_n$ and ${\cal H}_m$. Focus attention now in the  density
matrix $\rho \equiv \rho_{AB}$ for the associated composite system. If
\begin{enumerate}
\item $\rho$ has PPT, and
\item its rank ${\cal R}$ is such that  ${\cal R} \le {\rm max}(n,m)$,
\end{enumerate}
then, as was proved in \cite{HLVC00}, $\rho$ is separable. The above referred
to entropic criteria are also functional separability ones. Still another one
is majorization.

Let $\{\lambda_i\}$ be the set of eigenvalues of the matrix $\xi_1$
 and $\{\gamma_i\}$ be the set of eigenvalues of the matrix
 $\xi_2$. We assert that the ordered set of eigenvalues  
$\vec \lambda$ of $\xi_1$ {\it majorizes} the ordered set of 
eigenvalues  $\vec \gamma$ of $\xi_2$
(and writes $\vec \lambda \succ \vec \gamma$) when
$\sum_{i=1}^k \,\lambda_i \,\ge\, \sum_{i=1}^k \,\gamma_i$
for all $k$. It has been shown \cite{NK01} that, for all separable
 states $\rho_{AB} \equiv \rho$,

 \ben \label{NK} \vec \lambda_{\rho_A} &\succ&   \vec \lambda_{\rho}, \,\,
{\rm and}
 \nn
 \\
  \vec \lambda_{\rho_B} &\succ&   \vec \lambda_{\rho}.    \een
There is an intimate relation between this majorization criterion and entropic
inequalities, as discussed in \cite{T02,VW02}.

\section{Separability probabilities: exploring the whole state space}

We promised in the Introduction to perform a systematic numerical  survey of
the properties of arbitrary (pure and mixed) states of a given quantum system
by recourse to an exhaustive exploration of the concomitant  state-space ${\cal
S}$. To such an end it is necessary to introduce an appropriate measure $\mu $
on this space. Such a measure is needed to compute volumes within ${\cal S}$,
as well as to determine what is to be understood by a uniform distribution of
states on ${\cal S}$.  The natural measure that we are going to adopt here is
taken from the work of Zyczkowski {\it et al.} \cite{ZHS98,Z99}.
An arbitrary (pure or mixed) state $\rho$ of a quantum system
described by an $N$-dimensional Hilbert space can always be
expressed as the product of three matrices,

\be \label{udot} \rho \, = \, U D[\{\lambda_i\}] U^{\dagger}. \ee

\noindent Here $U$ is an $N\times N$ unitary matrix and
$D[\{\lambda_i\}]$ is an $N\times N$ diagonal matrix whose
diagonal elements are $\{\lambda_1, \ldots, \lambda_N \}$, with $0
\le \lambda_i \le 1$, and $\sum_i \lambda_i = 1$.
The group of unitary matrices $U(N)$ is
endowed with a unique, uniform measure: the Haar measure $\nu$
\cite{PZK98}. On the other hand, the $N$-simplex $\Delta$,
consisting of all the real $N$-uples $\{\lambda_1, \ldots,
\lambda_N \}$ appearing in (\ref{udot}), is a subset of a
$(N-1)$-dimensional hyperplane of ${\cal R}^N$. Consequently, the
standard normalized Lebesgue measure ${\cal L}_{N-1}$ on ${\cal
R}^{N-1}$ provides a natural measure for $\Delta$. The
aforementioned measures on $U(N)$ and $\Delta$ lead then to a
natural measure $\mu $ on the set ${\cal S}$ of all the states of
our quantum system \cite{ZHS98,Z99,PZK98}, namely,

\be \label{memu}
 \mu = \nu \times {\cal L}_{N-1}.
 \ee

 \noindent
  All our present considerations are based on the assumption
 that the uniform distribution of states of a quantum system
 is the one determined by the measure (\ref{memu}). Thus, in our
 numerical computations we are going to randomly generate
 states according to the measure (\ref{memu}).

\section{Survey's results}

\subsection{The overall scenario} 

An overall picture of the situation we encounter is sketched in Fig. 1, that is
to be compared to Fig. 3 of \cite{T02}. {\it Notice that our numerical
exploration allows us to dispense with Terhal's interrogation signs}. This
constitutes part of the original content of the present Communication.

 The set of all mixed states presents an onion-like shape, as conjectured by
 Terhal \cite{T02}.
 Which among these states are separable? As reviewed above, several criteria are
available. We start with the $q$-entropic one 
(see the Appendix and \cite{batle}).
 By using a definite value of $q$, namely $q = \infty$, and the sign of the
 associated, conditional $q$-entropy, we
 are able to define a
closed sub-region, whose states are supposedly separable. This region has a
definite border, that separates it from the sub-region of states entangled
according to this criterion. What we see now is that, if we use now {\it other}
separability-criteria, the associated sub-regions shrink in a manner prescribed
by the particular  criterion one employs. The shrinking process ends when one
reaches the sub-region defined by the Positive Partial Transpose (PPT)
criterion, which is a necessary and sufficient separability condition for $2
\times 2$ and $2 \times 3$ systems, being only necessary for higher dimensions.

Summing up, the volume of states which are separable according to different
criteria diminish as we use stronger and stronger criteria. There is a first
shrinking stage associated to entropic criteria, from its Von Neumann ($q=1$)
size, as $q$ grows,  to the limit case $q\rightarrow \infty$ \cite{batle}. A
second stage involves majorization, reduction, and, finally positive partial
transpose (PPT) \cite{T02}.

\subsection{PPT and Reduction}

We report now on our state-space exploration with regards to the probability of
finding a state with positive partial transpose. The results are depicted in
Fig. 2. The solid line corresponds to states with dimension $N=2 \times N_2$,
while the dashed line corresponds to $N=3 \times N_2$ states. Note how similar
are the pertinent  values in both cases. The tiny difference between them can
be inspected in the inset (a semi-logarithmic plot). To a good approximation,
our PPT probabilities decrease exponentially.

Fig. 3 deals instead with the probability of finding a state which obeys the
strictures of the reduction criterion, for $N=2 \times N_2$ (solid line) and
$N=3 \times N_2$ (dashed line). As a matter of fact, PPT and reduction coincide
for $N=2 \times N_2$. It is known that if a state satisfies PPT, it
automatically verifies the reduction criterion \cite{T02}. Here we have
demonstrated that, at least in the  $N=2 \times N_2$-instance, the converse is
also true. However, in the $N=3 \times N_2$-case, it is much more likely to
encounter a state that verifies reduction than one that verifies PPT.

\subsection{Entropic criteria and Majorization}

We begin with a brief recapitulation of former $q$-entropic results.
 The situation encountered in \cite{previous} was that the ``best" result
within the framework of the ``classical $q$-entropic inequalities" as a
separability criterion was reached using the limit case $q\rightarrow \infty$,
but considerably less attention was paid to other values of $q$. This was
remedied in \cite{batle}, where the question of $q$-entropic inequalities for
finite $q$-values was extensively discussed. It was there re-confirmed that the
above mentioned limit case does indeed the better job as far as separability
questions are concerned \cite{batle}. For such a reason, this limit $q$-value
is the only one to be employed below. See the Appendix for more details on
$q$-entropies.

In Fig. 4 we depict the probability of finding a state which, for $q
\rightarrow \infty$, has its two relative $q$-entropies positive (dashed
curves). In view of the intimate relation of entropic inequalities with
majorization \cite{T02,VW02}, we also analyze in Fig. 4  the probability that a
state is completely majorized by both of their subsystems (solid line).
 It is shown in \cite{VW02} that, if $\rho_{AB}$ satisfies the reduction 
criterion, its two associated relative $q$- entropies are non-negative as well.

 In the same work
 the authors assert that majorization is not implied by the relative entropy
 criteria. Our results confirm this assessment. In Fig. 4,
 the lower curves correspond to states $\rho$ with $N=2 \times N_2$,
while the upper curves have $N=3 \times N_2$. Majorization results and
$q$-entropic do coincide for two-qubits systems ($N_1=N_2=2$). More generally,
majorization probabilities  are a lower bound to probabilities for relative
$q$-entropic positivity, an interesting new result, as far as we know. Notice
also that the two approaches yield quite similar results in the 
$N=3 \times N_2$ case.

\subsection{Comparing more than two criteria together}

We compare now the reduction criterion to the PPT one. The former is implied by
the latter but is nonetheless a significant condition since its violation
implies the possibility of recovering entanglement by distillation, which is as
yet unclear for states that violate PPT \cite{VW02}. Fig. 5 a) depicts the
probability that state $\rho$ with $N=3 \times N_2$ either:
\begin{enumerate}
\item has a positive partial transpose and does not violate the reduction
criterion, or \item has a non positive partial transpose and violates
reduction.
\end{enumerate}
Remember that in the case $N=2 \times N_2$, the two criteria always coincide
\cite{T02}. For  $3 \times N_2$ the agreement between the two criteria becomes
better and better as $N_2$ augments.

Of more interest is to compare the relations among PPT, majorization, and the
entropic criteria (Fig. 5b), since it is not yet known how the majorization
criterion is related to other separability criteria like PPT, undistillability,
and reduction \cite{VW02}. In this vein, Fig. 5 b) plots the
``coincidence-probability" between, respectively,
\begin{enumerate}
\item  PPT and majorization (solid line), and
\item PPT and
the $q$-entropic criterion (dashed line).
\end{enumerate}
 The  curves on the top correspond to  $N=2 \times N_2$, 
while those at the bottom to
  $N=3 \times N_2$. In this last case the two curves agree with each other
  quite well.

  The conclusion here is that, as $N_2$ augments, the probability of coincidence
among the three criteria, and in particular  between majorization
  and PPT (our main concern),
   rapidly diminishes at first, and stabilizes
  itself afterwards. For two qubits the three criteria do agree with each other
  to a large extent.

 Fig. 6 a) depicts the probability that, for a given  state $\rho$,
\begin {enumerate}
\item reduction and majorization (solid line) and
\item reduction and the $q$-entropic criterion (dashed line)
\end{enumerate}
yield the same conclusion as regards separability. Without PPT in the game, and
opposite to what we encountered in Fig. 5, we find better coincidence for $N=3
\times N_2$ systems (top) than for $N=2 \times N_2$ (bottom). The deterioration
of the degree of agreement as $N_2$ grows is similar to that of Fig. 5, though.

Fig. 6 b) represents the probability that a state, for $q \rightarrow \infty$,
either:
\begin{enumerate}
\item has both positive relative $q$-entropies and satisfies the majorization 
criterion, or
\item has a negative relative $q$-entropy and is majorized by both of their
subsystems.
\end{enumerate}
The solid line corresponds to the case $N=2 \times N_2$,  while the dashed
lines corresponds to the $N=3 \times N_2$ instance. These results together with
those of  Figs. 4-5 could be read as implying   that majorization and the
$q$-entropic criteria provide almost the same answer for dimensions greater or
equal than $N=3 \times N_2$.

Finally,  in Fig. 7 we look for the  probability $P_{agree}$ that all criteria
considered in the present work do lead to the same conclusion on the
separability issue. $P_{agree}$ is plotted  as a function of the total
dimension $N=N_1 \times N_2$, with $N_1=2$ (solid line) and $N_1=3$ (dashed
line). The agreement is quite good for two qubits, deteriorates first as $N_2$
grows, and rapidly stabilizes itself around a value of 0.26 for $N_1=2$ and of
0.1 for $N_1=3$.

\subsection{Distilling}

Let us at now consider the results plotted in Fig. 8. We ask first  for the
relative number of states that violate the reduction criterion  and are thus
distillable \cite{Horo97} (solid line), and appreciate the fact that, as $N$
grows, so does the probability of finding distillable states.
 On the  other hand, the probability of encountering states that violate 
the majorization
criterion, represented by dashed lines, is much lower than that associated to
distillation.

For both criteria, the upper solid line corresponds to the case $N=2 \times
N_2$, and the lower one to $N=3 \times N_2$. The dashed curve with crosses
represents the case $N=2 \times N_2$, while the one with squares indicates the
$N=3 \times N_2$ instance. The dependence with $N_2$ of the dashed curves
(majorization violation) is not so strong as that of the solid ones
(distillability). Our results are lower bounds to the total volume of states 
that can be destilled.

\section{Conclusions}

We have performed a systematic numerical survey of the space of pure and mixed
states of bipartite systems of dimension $2 \times N_2$ and $3 \times N_2 $ in
order to investigate the relationships ensuing among different separability
criteria. Our main results are

\begin{itemize}
\item Regarding the  line of separability implication, see our graph in Fig. 1
and compare with the similar one of Terhal's (her Fig. 3). Her interrogation
signs receive an answer in our Fig.1
\item It is known that if a state satisfies PPT, it
automatically verifies the reduction criterion \cite{T02}. In the present work
we show that in the  $N=2 \times N_2$-instance, the converse is also true. In
the $N=3 \times N_2$-case, it is much more likely to encounter a state that
verifies reduction than one that verifies PPT.
\item We have numerically verified the assertion made in \cite{VW02} that
majorization is not implied by the relative entropic criteria. Majorization
results and $q$-entropic criteria coincide for two-qubits systems. In general,
majorization probabilities constitutes lower bound for relative $q$-entropic
positivity.
\item Regarding the relation between majorization and PPT, the agreement between
the criteria deteriorates as $N_2$ grows.

\item For dimensions  $\ge 3 \times N_2$, as illustrated by Figs. 4-5,
majorization and the $q$-entropic criteria
provide almost the same answers.
\end{itemize}

The present authors believe that the results of this numerical exploration shed
some light on the intricacies of the separability issue.

 \acknowledgments This work was partially supported by
the  MCyT grants BFM2002-0341 and SAB2001-006(Spain), by the Gouvernment of
Balearic Islandsand by CONICET (Argentine
Agency).

\newpage

\noindent {\bf FIGURE CAPTIONS}


\vskip 0.5cm

\noindent Fig. 1 - Schematics of the inclusion relations among separbility
criteria as given by the volume occupied by states $\rho$ for a given dimension
$N$ which fulfill them.\vskip 0.5cm

\noindent Fig. 2 - Probability of finding a state with positive partial
transpose. The solid line corresponds to states with dimension $N=2 \times
N_2$, while the dashed line corresponds to $N=3 \times N_2$ states.  The
difference between these curves can be appreciated in the inset 
(semi-logarithmic plot). Our probabilities  decrease, to a good approximation,
in exponential fashion.

\vskip 0.5cm

\noindent Fig. 3 - Probability of finding a state fulfilling the reduction
criterion for $N=2 \times N_2$ (solid line) and $N=3 \times N_2$ (dashed line).
The two probabilities coincide for $N=2 \times N_2$.

\vskip 0.5cm

\noindent Fig. 4 - Probability of finding a state whose two relative
$q$-entropies are positive for $q \rightarrow \infty$ (dashed curves). The
probability that a state be completely majorized by both of their subsystems is
represented by the solid line. Bottom: curves correspond to states $\rho$ with
$N=2 \times N_2$. Top: $N=3 \times N_2$. \vskip 0.5cm

\noindent Fig. 5 a) Probability that the state $\rho$ with $N=3 \times N_2$ 
either has i) a positive partial transpose {\it and} does not violate the 
reduction criterion, or ii) has a non positive partial transpose {\it and} 
violates reduction. In the case $N=2 \times N_2$ the outcome is always unity.
Fig. 5 b) Probability that \begin{enumerate} \item   PPT and majorization
(solid line) and, \item  PPT and the $q$-entropic criterion (dashed line)
\end{enumerate}lead to the same conclusion regarding separability.
 Top: $N=2 \times N_2$. Bottom:  $N=3 \times N_2$.

\vskip 0.5cm

\noindent  Fig. 6 a) Probability that
reduction and majorization (solid line) and reduction and the $q$-entropic
criterion (dashed line) yield the same conclusion reagrding separability. Top:
$N=3 \times N_2$. Bottom:  $N=2 \times N_2$ (lower curves).
Fig. 6 b) Probability that a state, for $q \rightarrow \infty$, either i) has
both positive relative $q$-entropies and fulfills majorization, or ii) has a
negative relative $q$-entropy and is majorized by both of their subsystems. The
solid line corresponds to the case $N=2 \times N_2$, while the dashed line
corresponds to $N=3 \times N_2$. \vskip 0.5cm

\noindent Fig. 7 - Total probability that all criteria considered in the present
work lead to  the same conclusion regarding separability. Probabilities are
plotted as a function of the total dimension $N=N_1 \times N_2$, with $N_1=2$
(solid line) and $N_1=3$ (dashed line). \vskip 0.5cm

\noindent Fig. 8 - Solid line: probability that a state  violates the reduction
criterion. Dashed line: the same for violation of the majorization criterion.
Top:  $N=2 \times N_2$. Bottom:  $N=3 \times N_2$. The dashed curve with
crosses represents the case $N=2 \times N_2$, while the one with squares
indicates the $N=3 \times N_2$ instance.
 \appendix
 \section{q-Information measures and the issue of quantum separability}

 There are several useful  entropic (or information) measures
   for the investigation of a quite important subject:  
the violation of classical entropic
  inequalities by quantum entangled states. The von Neumann measure

  \be \label{slog}
  S_1 \, = \,- \, Tr \left( \rho \ln \rho \right),
  \ee

  \noindent
  is important because of its relationship with the thermodynamic
  entropy. The $q$-entropy, which is a  function of the quantity

  \be \label{trq}
  \omega_q \, = \, Tr \left( \rho^q \right),
  \ee

  \noindent
  provides one with a whole family of entropic measures.
  In the limit $q\rightarrow 1 $ these measures incorporate (\ref{slog})
  as a particular instance. Most of the
  applications of $q$-entropies to physics involve either the R\'enyi entropy
  \cite{BS93},

  \be \label{renyi}
   S^{(R)}_q \, = \, \frac{1}{1-q} \, \ln \left( \omega_q \right),
  \ee

  \noindent
  or the Tsallis entropy \cite{T88,LV98,LSP01}

  \be \label{tsallis}
  S^{(T)}_q \, = \, \frac{1}{q-1}\bigl(1-\omega_q \bigr).
  \ee

  \noindent We reiterate that the von Neumann measure (\ref{slog}) constitutes a
particular instance of both R\'enyi's and Tsallis' entropies,  obtained in the
  limit $q\rightarrow 1$. The most distinctive single property of
  Tsallis' entropy is its nonextensivity. The Tsallis entropy of a
  composite system $A \oplus B $ whose state is described
  by a factorizable density matrix, $\rho_{AB} = \rho_A \otimes
  \rho_B$, is given by Tsallis' $q$-additivity law,

  \be
  S_q^{(T)}(\rho_{AB}) \, = \, S_q^{(T)}(\rho_A) \, + \, S_q^{(T)}(\rho_B)
  \, + \, (1-q)S_q^{(T)}(\rho_A)S_q^{(T)}(\rho_B).
   \ee

  \noindent
   In contrast, R\'enyi's entropy is extensive. That is,
   if $\rho_{AB} = \rho_A \otimes  \rho_B$,

  \be
  S_q^{(R)}(\rho_{AB}) \, = \, S_q^{(R)}(\rho_A) \, + \,
  S_q^{(R)}(\rho_B).
  \ee

  \noindent
  Tsallis' and R\'enyi's measures are
  related through

  \be \label{stsr}
  S^{(T)}_q \, = \,F( S^{(R)}_q),
  \ee

  \noindent
  where the function $F$ is given by

  \be \label{fx}
  F(x) \, = \, \frac{1}{1-q} \left\{ e^{(1-q)x} - 1 \right\}.
  \ee

  \noindent
  An immediate consequence of equations (\ref{stsr}-\ref{fx})
  is that, for all non vanishing values of $q$, Tsallis' measure
  $ S^{(T)}_q$ is a monotonic increasing function of R\'enyi's
  measure $ S^{(R)}_q $.

  Considerably attention has been recently paid to a relative
  entropic measure based upon Tsallis' functional defined as

  \be \label{qurela}
  S^{(T)}_q(A|B) \, = \,
  \frac{S^{(T)}_q(\rho_{AB})-S^{(T)}_q(\rho_B)}{1+(1-q)S^{(T)}_q(\rho_B)}.
  \ee

  \noindent
  Here $\rho_{AB}$ designs an arbitrary quantum state of the
  composite system $A\oplus B$,
  not necessarily factorizable nor separable,
  and $\rho_B = Tr_A (\rho_{AB})$. The relative $q$-entropy
  $S^{(T)}_q(B|A)$ is defined in a similar way as (\ref{qurela}),
  replacing $\rho_B $ by $\rho_A = Tr_B (\rho_{AB})$.
     The relative $q$-entropy
  (\ref{qurela}) has been recently studied in connection
  with the separability of density matrices describing composite
  quantum systems \cite{TLB01,TLP01}. For separable states,
  we have \cite{VW02}

  \ben \label{qsepar}
  S^{(T)}_q(A|B) &\ge & 0, \cr
  S^{(T)}_q(B|A) &\ge & 0.
  \een

  \noindent
  On the contrary, there are entangled states that have negative
  relative $q$-entropies. That is, for some entangled states one (or both)
  of the inequalities (\ref{qsepar}) are not verified.

  Notice that the denominator in (\ref{qurela}),

  \be \label{deno}
  1+(1-q)S^{(T)}_q \, = \, w_q \, > \, 0.
  \ee

  \noindent
  is always positive. Consequently, as far as the sign of the
  relative entropy is concerned, the denominator in (\ref{qurela})
  can be ignored. Besides, since Tsallis' entropy is a monotonous
  increasing function of R\'enyi's (see Equations (\ref{stsr}-\ref{fx})),
  it is plain that (\ref{qurela}) has always the same sign as

  \be \label{relarenyi}
  S^{(R)}_q(A|B) \, = \, S^{(R)}_q(\rho_{AB})-S^{(R)}_q(\rho_{B}).
  \ee

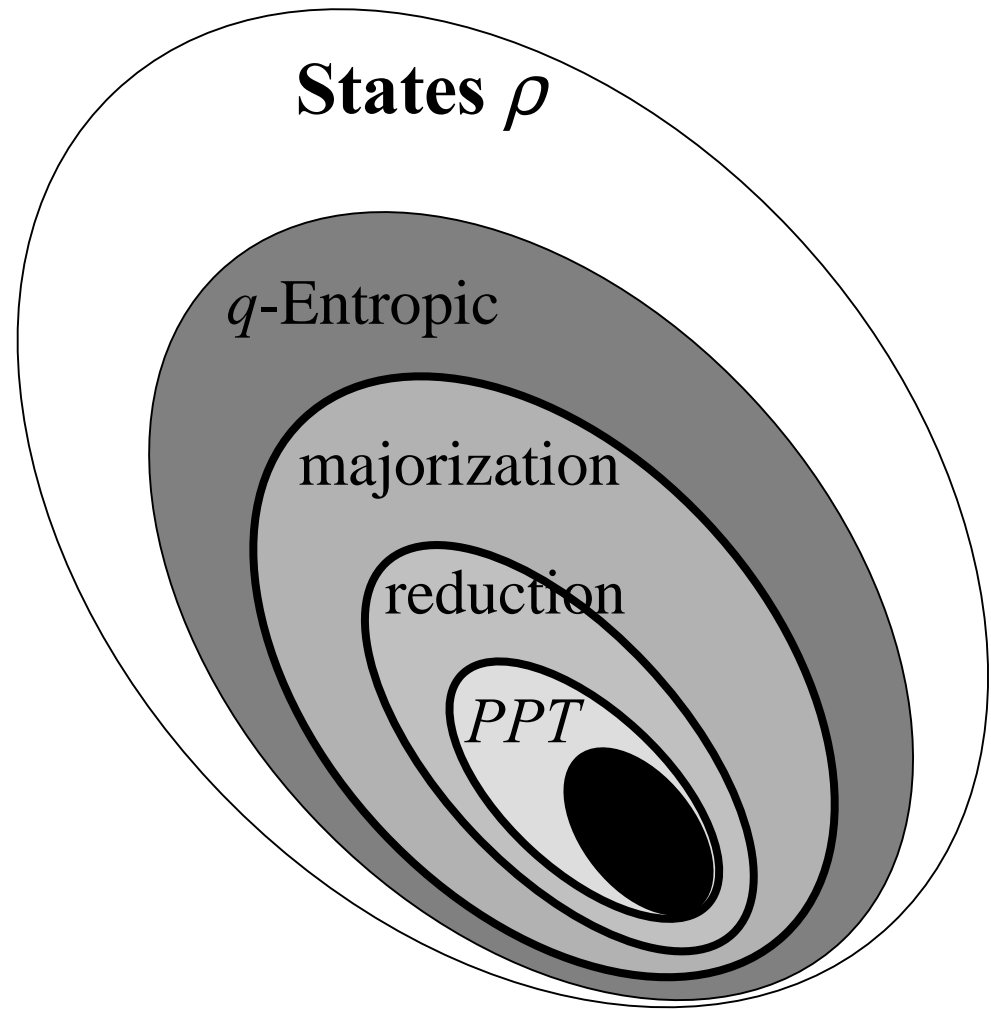
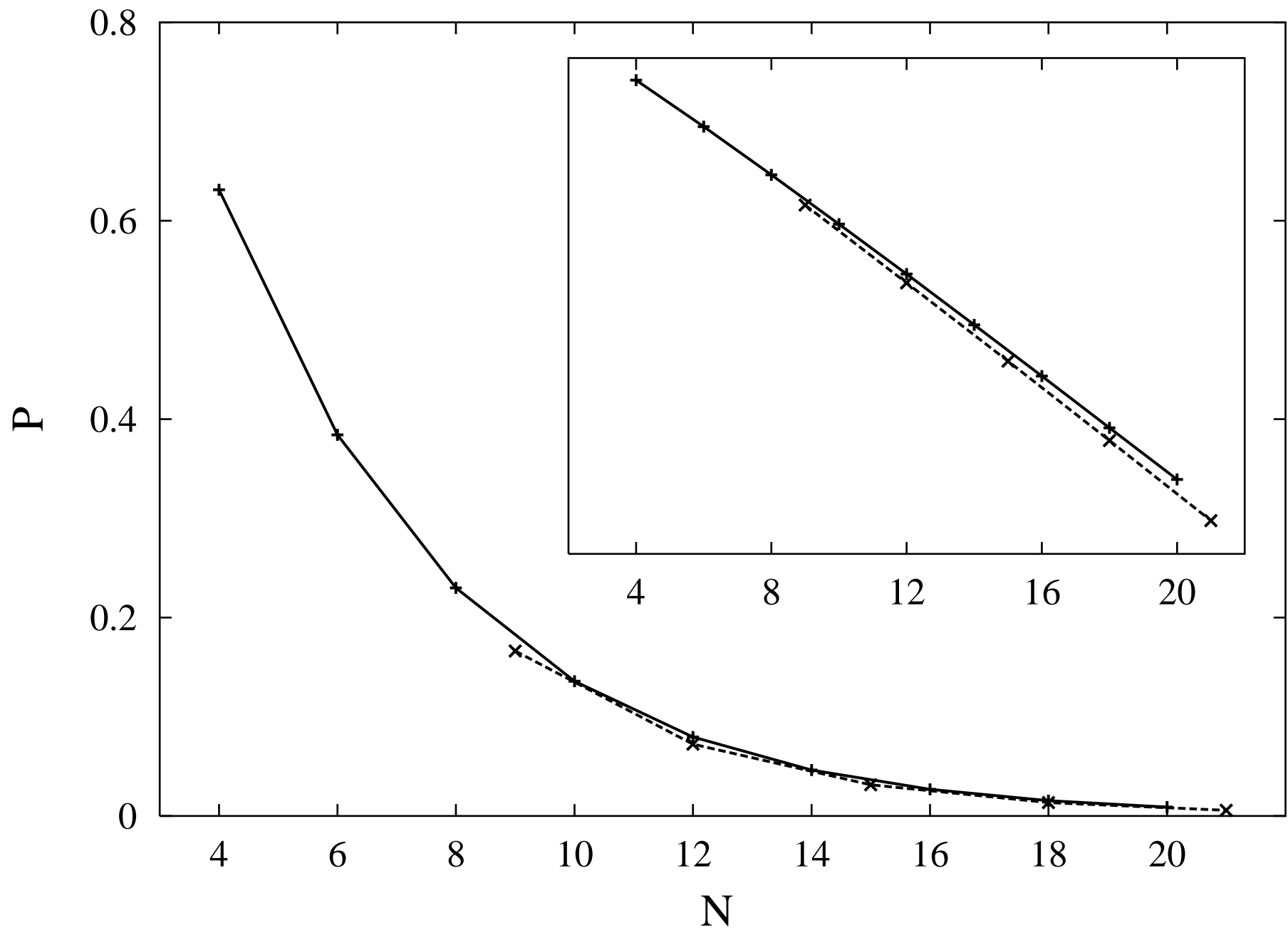
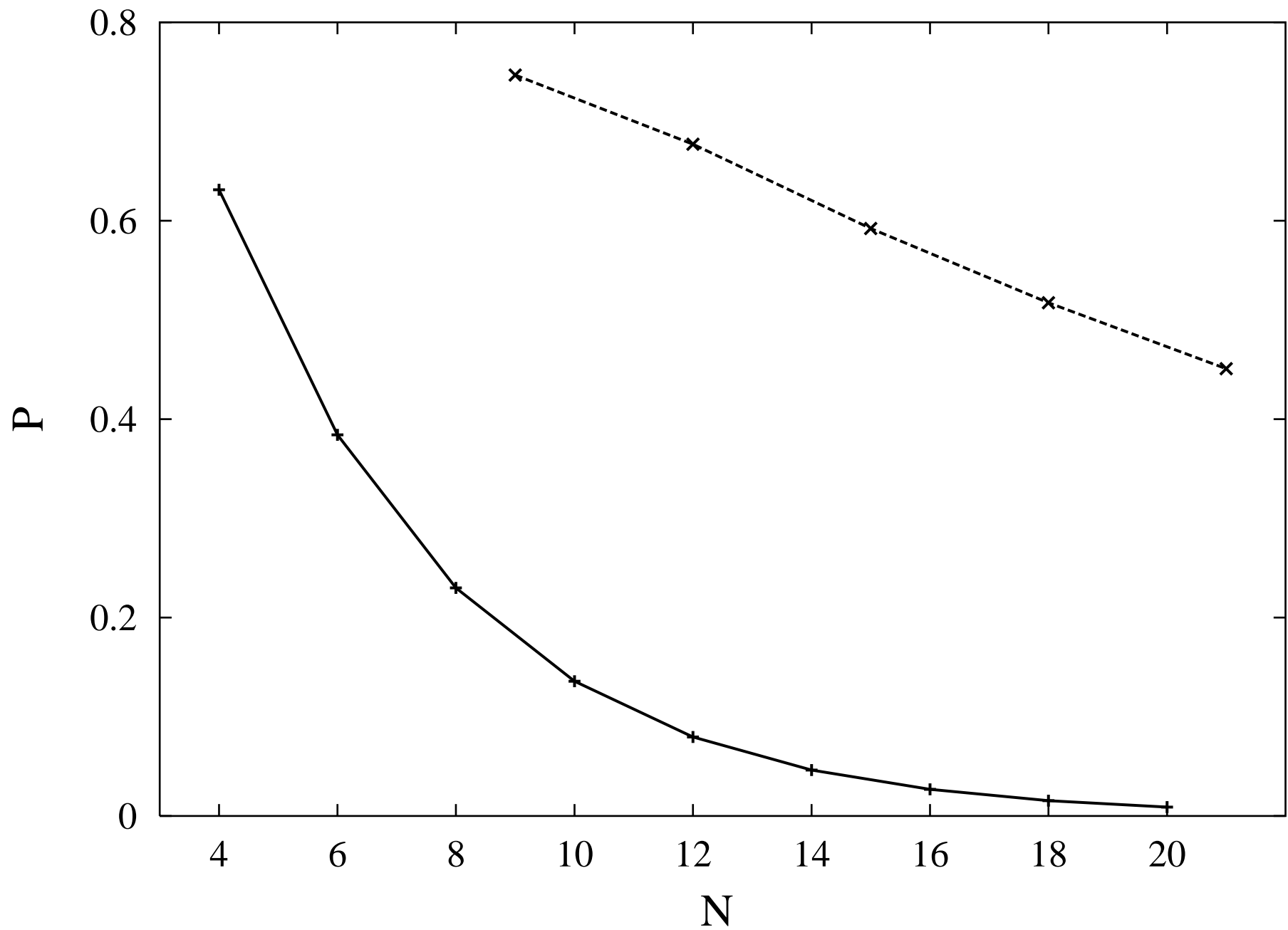
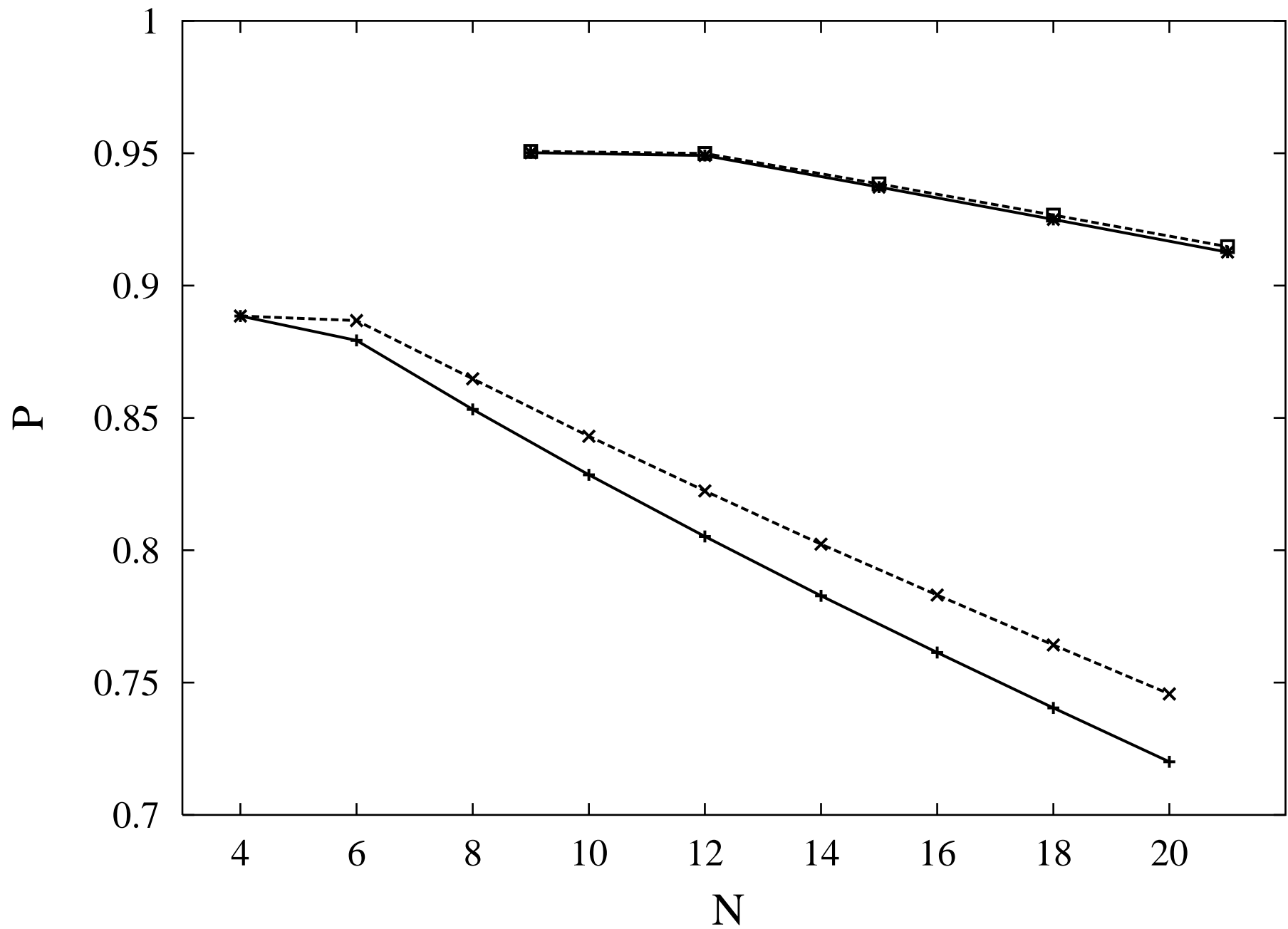
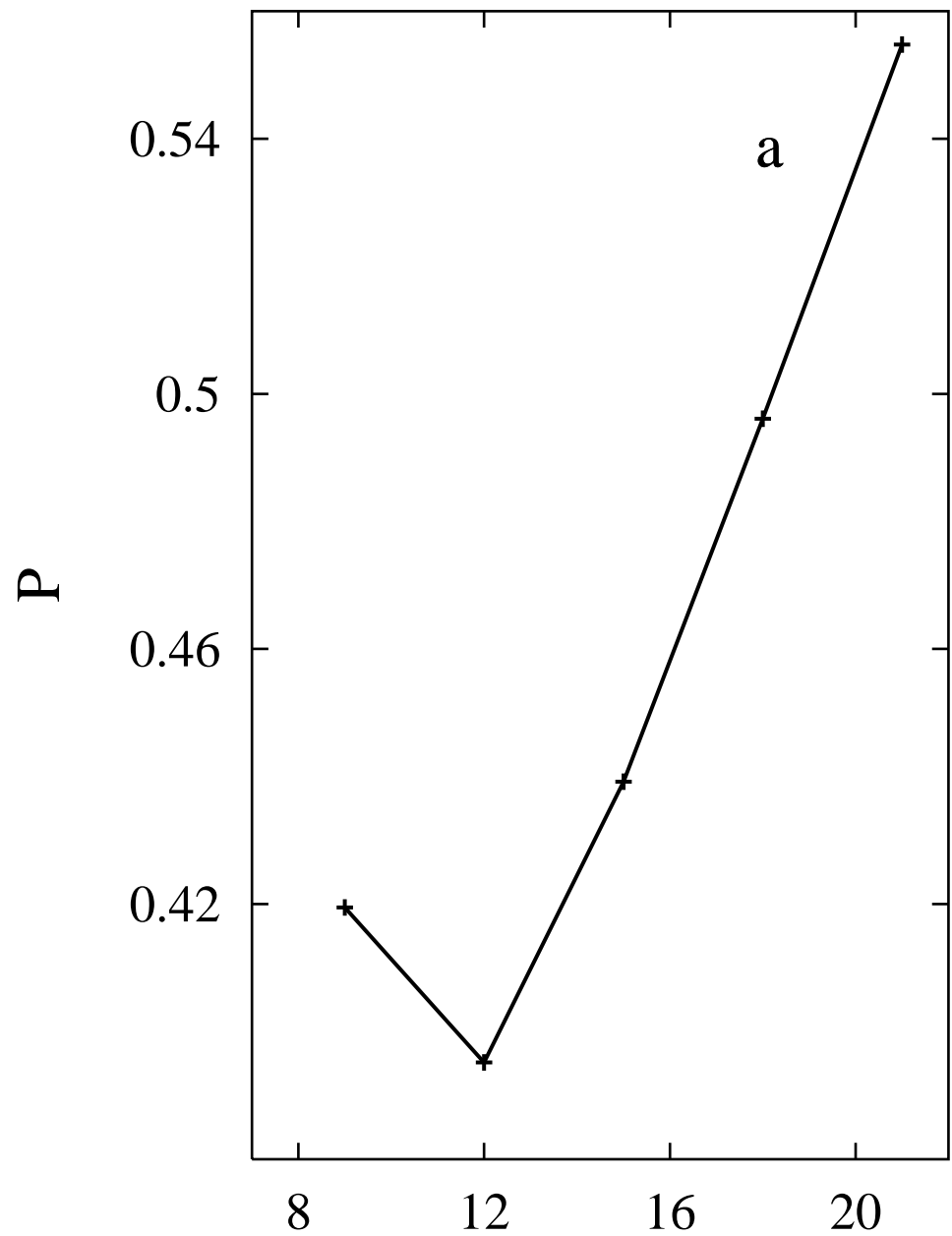
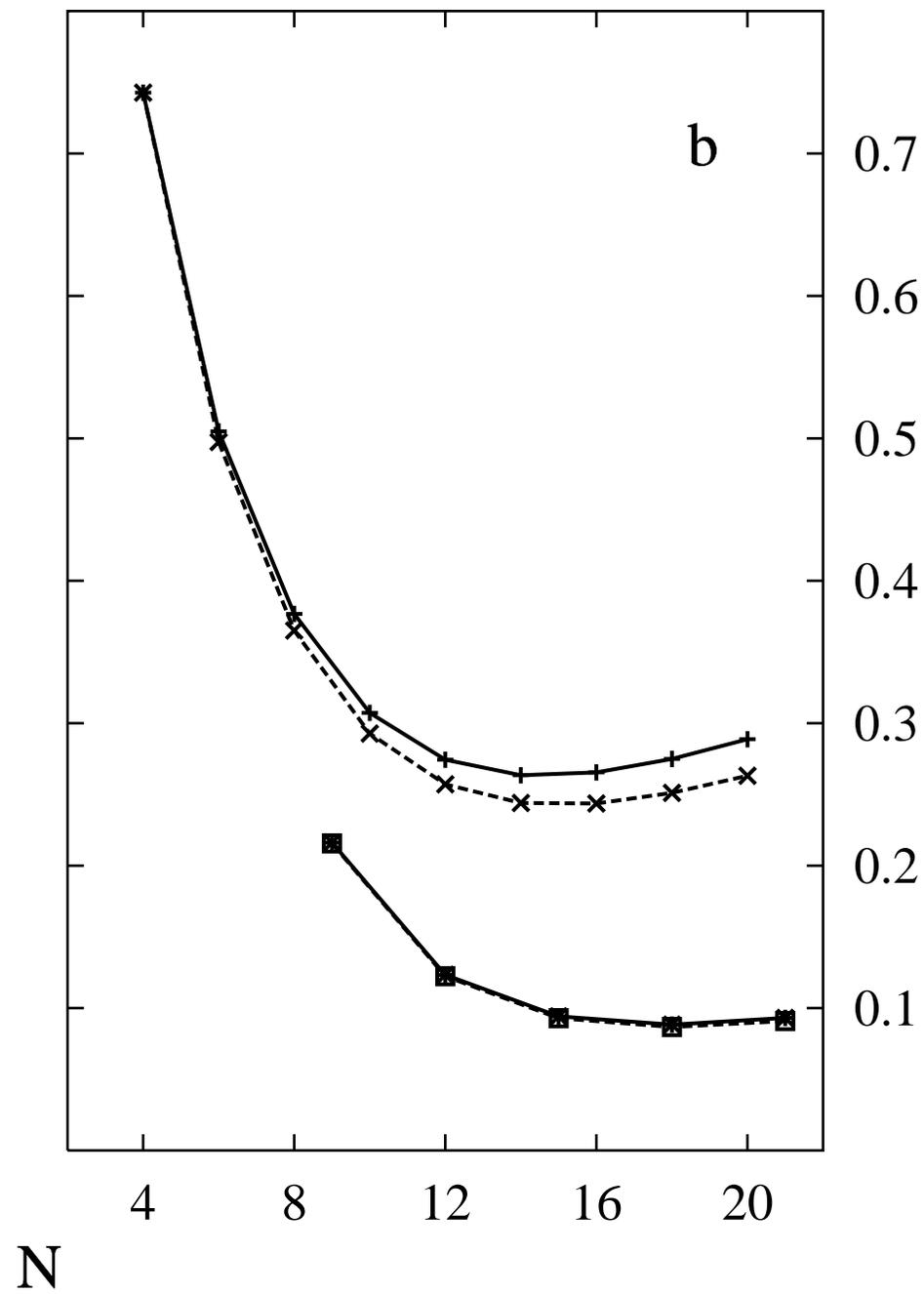
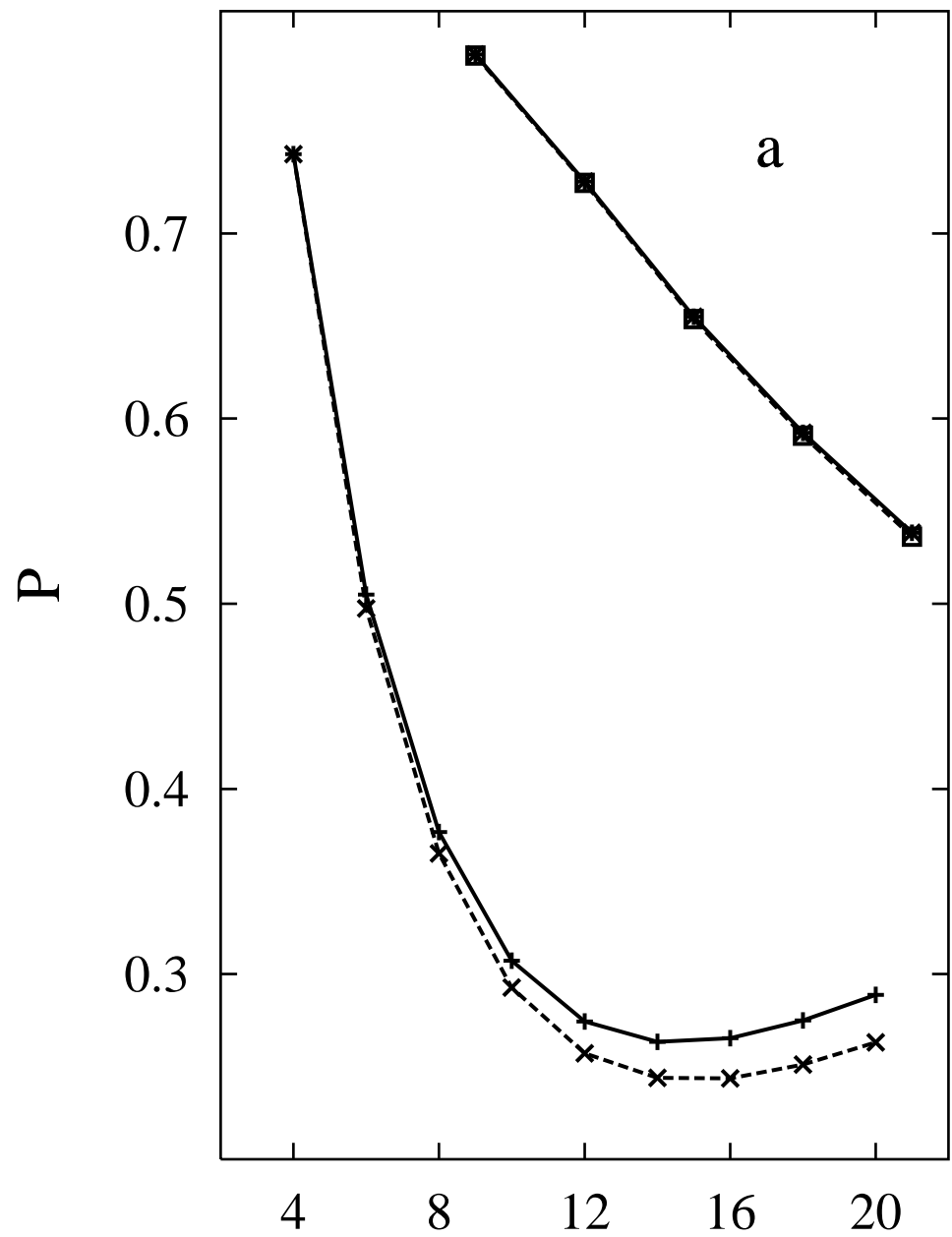
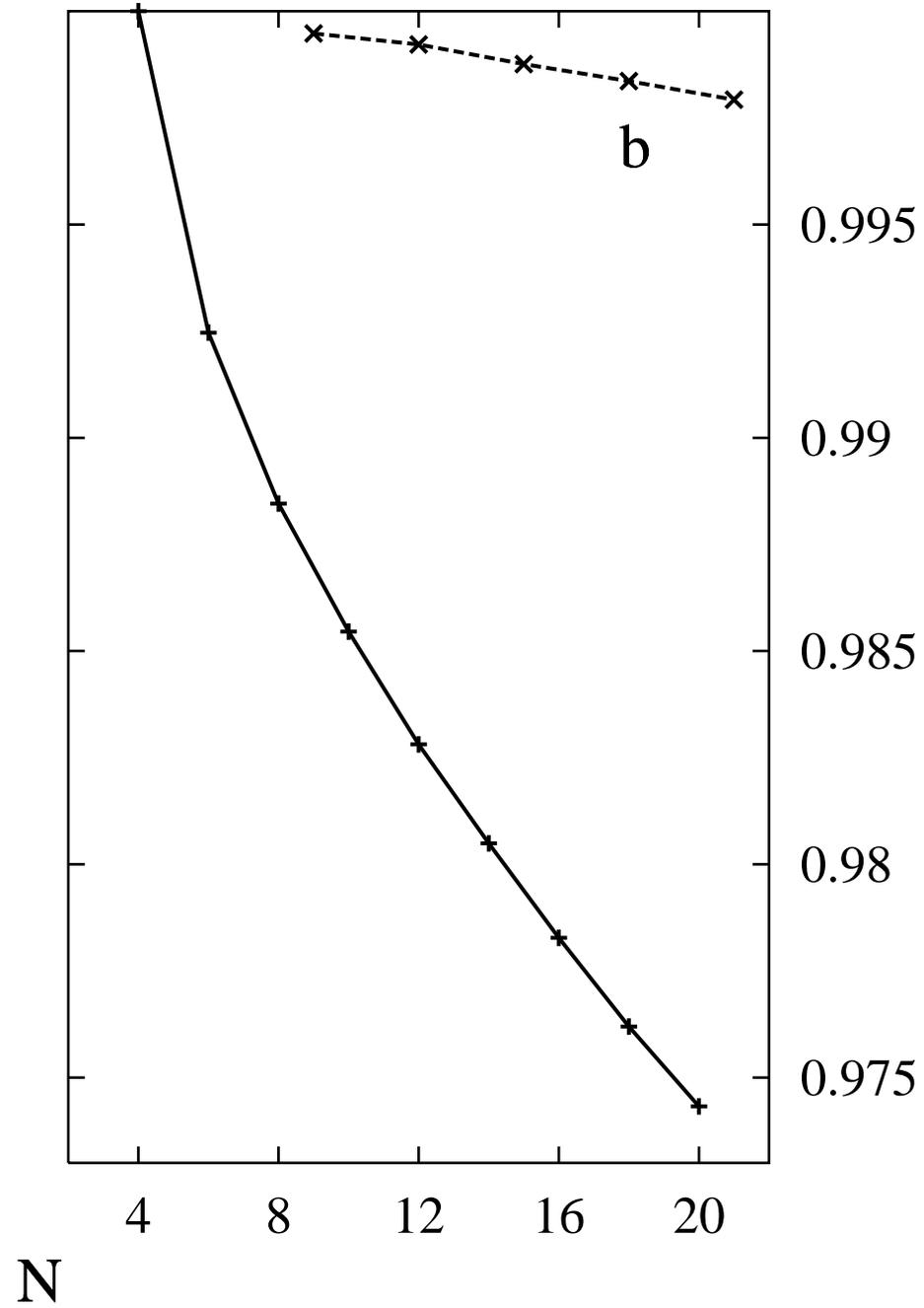
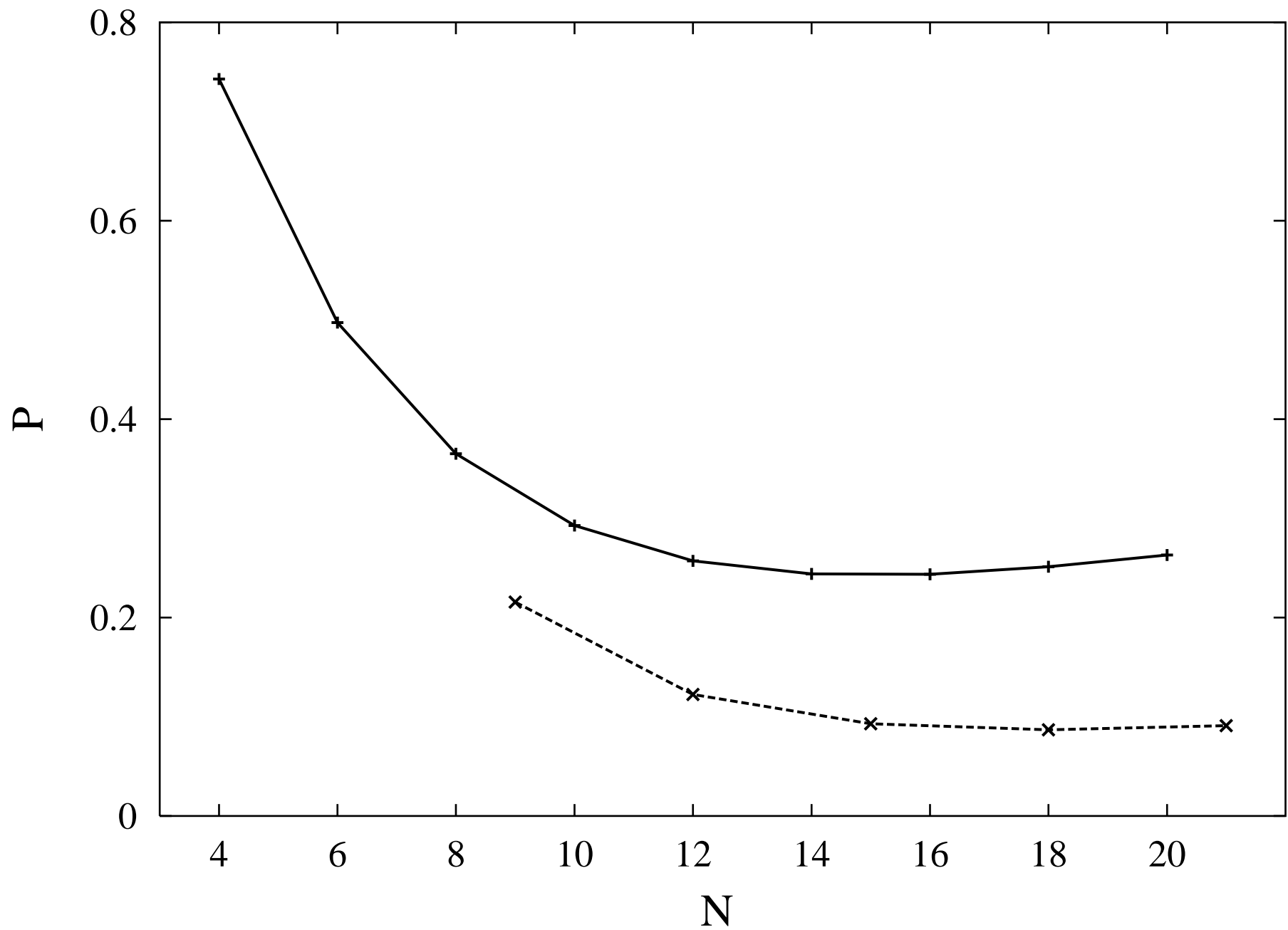
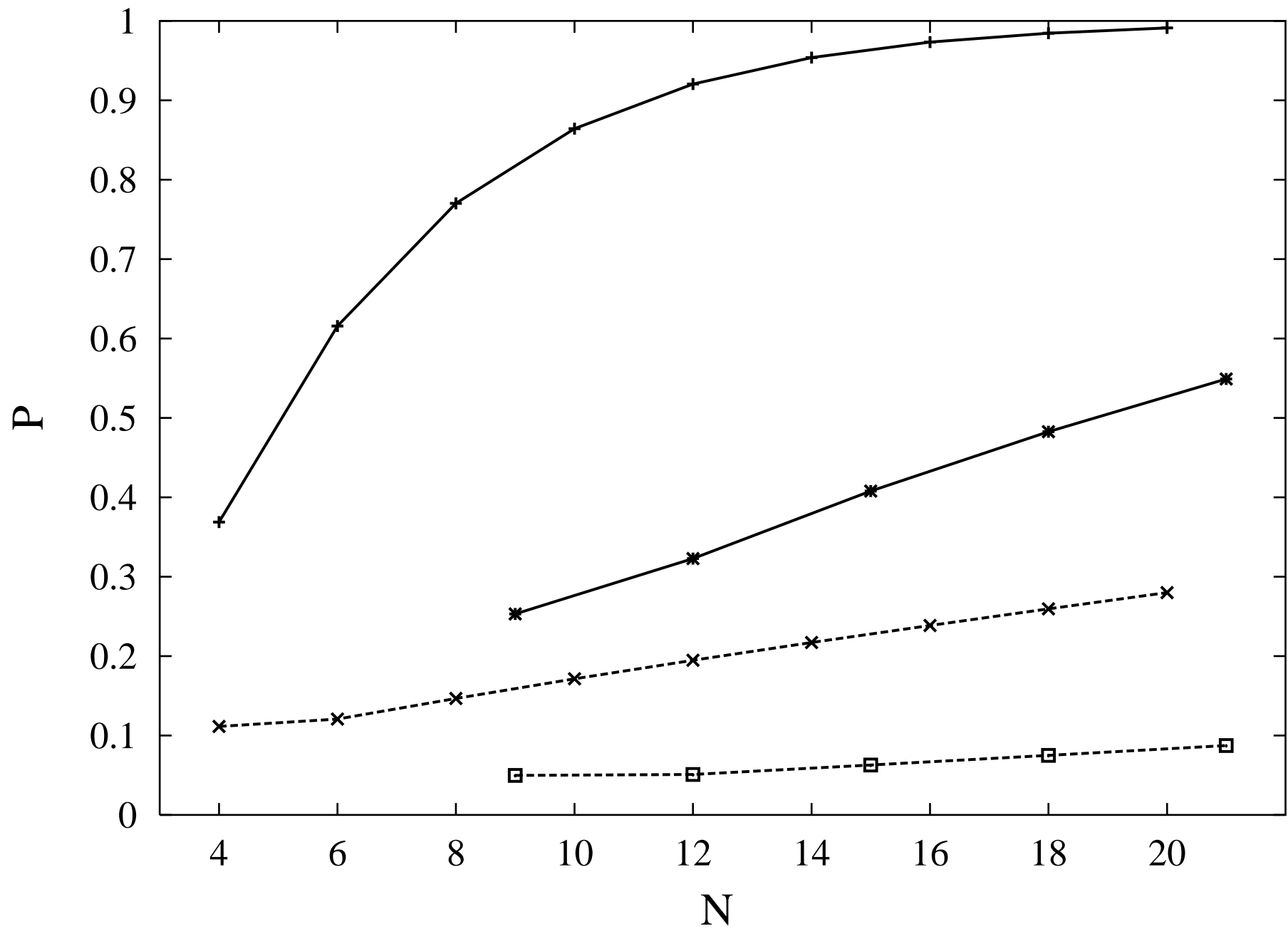
\end{document}